\documentstyle[multicol,aps,epsf]{revtex}

\title{Optimized Forest-Ruth- and Suzuki-like algorithms for integration \\
       of motion in many-body systems}

\author{I. P. Omelyan,$^{1,2}$ I. M. Mryglod,$^{1,2}$ and R. Folk$^2$}

\address{$^1$Institute for Condensed Matter Physics,
         1 Svientsitskii Street, UA-79011 Lviv, Ukraine}
\address{$^2$Institute for Theoretical Physics, Linz University,
         A-4040 Linz, Austria}

\date{\today}

\begin{document}

\maketitle

\begin{abstract}

An approach is proposed to improve the efficiency of fourth-order
algorithms for numerical integration of the equations of motion
in molecular dynamics simulations. The approach is based on
an extension of the decomposition scheme by introducing extra
evolution subpropagators. The extended set of parameters of the
integration is then determined by reducing the norm of truncation
terms to a minimum. In such a way, we derive new explicit symplectic
Forest-Ruth- and Suzuki-like integrators and present them in
time-reversible velocity and position forms. It is proven that
these optimized integrators lead to the best accuracy in the
calculations at the same computational cost among all possible
algorithms of the fourth order from a given decomposition class.
It is shown also that the Forest-Ruth-like algorithms, which are
based on direct decomposition of exponential propagators, provide
better optimization than their Suzuki-like counterparts which
represent compositions of second-order schemes. In particular,
using our optimized Forest-Ruth-like algorithms allows us to
increase the efficiency of the computations more than in ten
times with respect to that of the original integrator by Forest
and Ruth, and approximately in five times with respect to Suzuki's
approach. The theoretical predictions are confirmed in molecular
dynamics simulations of a Lennard-Jones fluid. A special case of
the optimization of the proposed Forest-Ruth-like algorithms to
celestial mechanics simulations is considered as well.

\vspace{6pt}
\noindent Pacs numbers: 02.60.Cb; 02.70.Ns; 05.10.-a;
45.50.Pk; 95.75.Pq

\vspace{6pt} \noindent Keywords: molecular dynamics, fourth-order
algorithms, decomposition scheme, optimized algorithms, celestial
mechanics simulations.

\end{abstract}

\vspace{13pt}

\begin{multicols}{2}

\section{Introduction}

Modelling various physical and chemical processes in molecular
dynamics (MD) simulations we come to the necessity to integrate
the equations of motion for a many-body system of interacting
particles. A lot of numerical algorithms have been devised and
implemented over the years to perform such an integration. The
traditional high-order explicit Runge-Kutta (RK) and implicit
predictor-corrector (PC) schemes \cite{Gear,Burden} were applied
in early investigations. Is was soon realized that the extra orders
obtained in these schemes are not relevant since the truncation
errors accumulate drastically on MD scales of time \cite{Allen}.
This high instability restricts the application of RK and PC
integrators in long-term MD simulations to very small time steps
only, and, thus, reduces significantly the efficiency of the
computations. In addition, the RK and PC algorithms produce
solutions which, unlike exact phase trajectories, are neither
symplectic nor time reversible.

In 1990, a new approach to the integration of motion in many-body systems
has been proposed \cite{Yoshida,Forest,Suzukip,Suzukium,Suzuki}. Within
this approach, the time propagation is carried out on the basis of
exponential decompositions of evolution propagators. The main advantage
of the decomposition method is that for an arbitrary order in the time
step it allows to construct algorithms which are exactly symplectic and
time-reversible. The preservation of symplecticity and reversibility
appears to be very important because, as is now well established, this
closely relates to the stability of an algorithm \cite{Frenkel}. Another
nice property of the decomposition integration is its explicitness and
simplicity in implementation. This is in a sharp contrast to implicit
time-reversible symplectic algorithms obtained recently \cite{Sofron}
within the RK approach, where cumbersome systems of coupled nonlinear
equations must be solved by iteration at each step of the integration
process.

Nowadays, the decomposition method should be considered as the main
tool for construction of efficient integrators of motion in classical
as well as quantum systems \cite{Suzuki}. Modified versions of this
method have also been introduced. In particular, it was shown that for
atomic systems with long-range interactions the efficiency of the
integration can be improved by additionally splitting the Liouville
operator into slow and fast components \cite{Tuckerman,Stuart}.
In such a multiple scale propagation, the slow subdynamics is
treated in a specific way using larger step sizes in view of
the weakness of long-range forces. The fasted motion, caused by
the interactions at short interparticle distances, remains to be
integrated with the help of original decomposition algorithms. The
question of how to derive higher-order integrators by composing
lower-order decomposition schemes has been considered as well
\cite{Suzukium,Qin,McLachlan}. Moreover, it has been shown recently
how to adapt the decomposition approach to integrate not only
translational motion in atomic liquids, but also simulate more
complicated molecular and spin systems at the presence of
orientational degrees of freedom \cite{Dullw,Krech,Omfes}.

The main attention in previous studies has been directed to obtain
algorithms which require a minimal number of force evaluations per
time step. For instance, the well-known velocity- and position-Verlet
integrators \cite{Tuckerman,Swope} correspond to a second-order
decomposition scheme with one force evaluation per step. The fourth-order
algorithm by Forest and Ruth \cite{Forest} presents a scheme with three
such force recalculations. Sixth-order schemes \cite{Yoshida,Suzukium}
can be reproduced beginning from seven evaluations of force for each
particle during each time step. It is worth mentioning that sixth- and
higher-order schemes involve too large number of force recalculations
and generally are not recommended to be used in MD simulations. The
reasons are that for a system with a great number of particles, the force
evaluations constitute the most time-consuming part of the computations
and that the truncations errors decrease with increasing the order of the
scheme much slower than the total number of these evaluations. Such
high-order algorithms may be efficient only for systems composed of
a few bodies when a very high precision in determination of phase
trajectories is desirable, for instance, in astronomical applications.
In the case of MD simulations, the preference should be given to
more simple second- or fourth-order schemes. For many MD applications
to achieve a required level of accuracy, the fourth-order integrators
appear to be more efficient than second-order schemes.

The minimal numbers of force evaluations per time step do not
guarantee, however, the optimization with respect to the overall
number of such evaluations which are necessary to perform during a
fixed observation time interval. This is so because schemes, in
which these numbers exceed the minimum, can be used with larger
sizes of the time step in order to obtain the same accuracy in
solutions. Only very few papers \cite{Suzukium,KahanLi,OmNew} were
devoted to derivation of such extended schemes. In particular,
Suzuki {\em et al.} have pointed out that the fourth-order
algorithm by Forest and Ruth may not lead to optimal performance
since during the evaluation it involves time coefficients which
are larger in magnitude than the size of the initial time step
\cite{Suzukium}. As a result, the original algorithm has been
replaced by a new fourth-order integrator composing five, instead
of three, second-order Verlet schemes. However, the question of
how this replacement influences on the efficiency of the
computation has not been studied. Extended Suzuki-like schemes
were also the subject of investigations by Kahan and Li
\cite{KahanLi}. Composing second-order algorithms, they introduced
higher-order integrators with time coefficients chosen to provide
a minimum for the maximal one among them in magnitude or for the
sum of absolute values of these coefficients. Again, it has been
unknown whether the increased number of force evaluations in their
integrators is compensated by the possibility of using larger step
sizes or not. In Ref.~\cite{OmNew} new explicit velocity- and
position-Verlet-like algorithms of the second order were proposed
to integrate the equations of motion in many-body systems. These
algorithms were derived on the basis of an extended decomposition
scheme at the presence of a free parameter. The nonzero value for
this parameter was obtained by reducing the influence of truncated
terms to a minimum. As a result, the new optimized algorithms
appear to be more efficient than the original Verlet versions
which correspond to a particular case when the introduced
parameter is equal to zero.

In the present paper we propose a consequent approach for
improving the efficiency of fourth-order decomposition
integration. As a result, we derive new Forest-Ruth- and
Suzuki-like integrators by explicitly reducing the influence of
truncation terms to a minimum. Although this requires (as in any
extended scheme) extra force evaluations per time step, the
overall number of force recalculations keeps unchanged due to the
use of larger step sizes. At the same time, the resulting
precision increases significantly with respect to that of the
original algorithms by Forest and Ruth, and Suzuki. It is
demonstrated also that previous criteria used for optimization of
decomposition algorithms may have nothing to do with providing the
best performance of the computations.

\section{Decomposition integration}

We will deal with a classical $N$-body system described by the Hamiltonian
\begin{equation}
H = \sum_{i=1}^N \frac{m {{\bf v}_i}^2}{2} +
    \frac12 \sum_{i \ne j}^N \varphi(r_{ij}) ,
\end{equation}
where ${\bf r}_i$ and ${\bf v}_i$ denote the position and velocity,
respectively, of particle $i$ with mass $m$, and $\varphi(r_{ij})$
is the interparticle potential of interaction with $r_{ij}=|{\bf r}_i-
{\bf r}_j|$. The equations of motion for such a system can be cast in
the following compact form
\begin{equation}
\frac{{\rm d} {\mbox{\boldmath $\rho$}}}{{\rm d} t} =
[{\mbox{\boldmath $\rho$}} \circ H]
\equiv
L {\mbox{\boldmath $\rho$}}(t) .
\end{equation}
Here ${\mbox{\boldmath $\rho$}} \equiv \{ {\bf r}_i, {\bf v}_i \}$
is the full set ($i=1,2,\ldots,N$) of phase variables, $[\,\circ\,]$
represents the Poisson bracket and
\begin{equation}
L = \sum_{i=1}^N \Big( {\bf v}_i {\mbox{\boldmath $\cdot$}}
\frac{\partial}{\partial {\bf r}_i}+\frac{{\bf f}_i}{m}
{\mbox{\boldmath $\cdot$}} \frac{\partial}{\partial {\bf v}_i}
\Big)
\end{equation}
is the Liouville operator
with ${\bf f}_i\!=\!-\sum_{j (j \ne i)}^N \varphi'(r_{ij})
{\bf r}_{ij}/r_{ij}$
being the force acting on particles due to the interaction.

If an initial configuration ${\mbox{\boldmath $\rho$}}(0)$ is
provided, the unique solution to Eq.~(2) can be presented as
\begin{equation}
{\mbox{\boldmath $\rho$}}(h)={\rm e}^{Lh}
{\mbox{\boldmath $\rho$}}(0) \equiv
{\rm e}^{(A+B)h} {\mbox{\boldmath $\rho$}}(0) \, ,
\end{equation}
where $h$ denotes the time step, and the Liouville operator $L=A+B$ has
been split into the free-motion $A={\bf v} {\mbox{\boldmath $\cdot$}}
\partial/\partial {\bf r}$ and potential $B={\bf f}/m {\mbox{\boldmath
$\cdot$}} \partial/\partial {\bf v}$ parts with ${\bf v} \equiv \{ {\bf
v}_i \}$, ${\bf r} \equiv \{ {\bf r}_i \}$, and ${\bf f} \equiv \{ {\bf
f}_i \}$. The significance of such a splitting will be understood below.
Then the evolution of the system can be investigated during arbitrary
times $t$ by repeating the single-step propagation, ${\mbox{\boldmath
$\rho$}}(t) = \big( {\rm e}^{Lh} \big)^l {\mbox{\boldmath $\rho$}}(0)
\equiv \big( {\rm e}^{(A+B) h} \big)^l {\mbox{\boldmath $\rho$}}(0)$,
where $l=t/h$ is the total number of steps.

Of course, solution (4) is quite formal because the exponential propagator
$e^{L h}$ does not allow to be evaluated exactly at any $h$ (solutions in
quadratures are possible only for $N=2$ that is not relevant for our case
of many-body systems when $N \gg 1$). However, at small enough values of
$h$, the total propagator can be decomposed \cite{Yoshida,Forest,Suzukip,%
Suzukium,Suzuki} using the formula
\begin{equation}
{\rm e}^{(A+B)h + {\cal O}(h^{K+1})}
= \prod_{p=1}^{P} {\rm e}^{A a_p h} {\rm e}^{B b_p h} .
\end{equation}
The coefficients $a_p$ and $b_p$ in this formula should be chosen in
such a way to provide the highest possible value for $K \ge 1$ at a
given integer number $P \ge 1$. Then integration (4) can be performed
approximately with the help of Eq.~(5) by neglecting truncation terms
${\cal O}(h^{K+1})$.

The main advantage of the above decomposition is that the exponential
subpropagators ${\rm e}^{A \tau}$ and ${\rm e}^{B \tau}$, appearing
in the right-hand-side of Eq.~(5), are analytically integrable. Indeed,
as can be shown readily,
\begin{eqnarray}
&&{\rm e}^{A \tau} {\mbox{\boldmath $\rho$}} \equiv
{\rm e}^{{\bf v} {\mbox{\boldmath $\cdot$}} \partial/\partial {\bf r}
\tau} \{ {\bf r}, {\bf v} \} = \{ {\bf r} + \tau {\bf v}, {\bf v} \} ,
\nonumber \\ [-5pt] \\ [-5pt]
&&{\rm e}^{B \tau} {\mbox{\boldmath $\rho$}} \equiv {\rm e}^{{\bf
f}/m {\mbox{\boldmath $\cdot$}} \partial/\partial {\bf v} \tau} \{
{\bf r}, {\bf v} \} = \{ {\bf r}, {\bf v} + \tau {\bf f}/m \} \nonumber
\end{eqnarray}
with $\tau$ being equal to $a_p h$ or $b_p h$. That is why we explicitly
presented $L$ as the sum of free-motion $A$ and potential $B$ terms.
Another important feature of the decomposition integration is that it
leads to symplectic trajectories. This is so since Eq.~(6) represents,
in fact, simple shifts of position and velocity, and these shifts do not
change the volume in phase space. The time reversibility $S(-t) {\mbox
{\boldmath $\rho$}}(t) = {\mbox{\boldmath $\rho$}}(0)$ of solutions
(following from the property $S^{-1}(t)=S(-t)$ of evolution operator
$S(t)={\rm e}^{Lt}$) can also be reproduced by imposing additional
constraints on the coefficients $a_p$ and $b_p$, namely, $a_1=0$,
$a_{p+1}=a_{P-p+1}$, and $b_p=b_{P-p+1}$, or $a_p=a_{P-p+1}$, and
$b_p=a_{P-p}$ at $b_P=0$. Then the subpropagators ${\rm e}^{A \tau}$
and ${\rm e}^{B \tau}$ will enter symmetrically in the decompositions
and, thus, provide automatically the required reversibility. Note also
that such constraints lead to automatic disappearing even-order terms
$\propto h^{2k}$ in the function ${\cal O}(h^{K+1})$ for any $k \ge 0$.
For this reason, the order $K$ of time-reversible algorithms may accept
only even numbers, ($K=2,4,6,\ldots$). The cancellation of odd-order
terms $\propto h^{2k+1}$ in ${\cal O}(h^{K+1})$ up to a required finite
number $k$ will be provided by fulfilling a set of basic conditions for
$a_k$ and $b_k$. For example, the condition $\sum_{p=1}^P a_p=\sum_{p=1}^P
b_p=1$ is necessary to cancel the first-order truncation uncertainties.

The decomposition method is quite general to build numerical integrators
of arbitrary orders. In particular, the second-order ($K=2$) velocity-Verlet
(VV) algorithm \cite{Tuckerman,Swope} is obtained from Eq.~(5) at $P=2$ and
$a_1=0$, $b_1=b_2=1/2$, $a_2=1$, i.e., ${\rm e}^{(A+B)h +{\cal O}(h^3)} =
{\rm e}^{B h/2} {\rm e}^{A h} {\rm e}^{B h/2}$. The case when the operators
$A$ and $B$ are replaced by each other ($A \leftrightarrow B$) is also
possible, and we come \cite{Tuckerman} to the position-Verlet (PV)
integrator
\begin{equation}
{\rm
e}^{(A+B)h +{\cal O}(h^3)} = {\rm e}^{A \frac{h}{2}} {\rm e}^{B h}
{\rm e}^{A \frac{h}{2}} \, ,
\end{equation}
corresponding to the choice $a_1=a_2=1/2$, $b_1=1$, and $b_2=0$. The
fourth-order ($K=4$) algorithm by Forest and Ruth (FR) \cite{Forest} is
immediately reproduced from Eq.~(4) at $P=4$
\begin{eqnarray}
{\rm e}^{(A+B) h + {\cal O}(h^5)} = &&
{\rm e}^{A \theta \frac{h}{2}} {\rm e}^{B \theta h}
{\rm e}^{A (1-\theta) \frac{h}{2}}
{\rm e}^{B (1-2\theta) h} \times
\nonumber \\ [-6pt] \\ [-6pt] &&
{\rm e}^{A (1-\theta) \frac{h}{2}} {\rm e}^{B \theta h}
{\rm e}^{A \theta \frac{h}{2}}
\nonumber , \ \ \ \ \
\end{eqnarray}
with $a_1=a_4=\theta/2$, $a_2=a_3=(1-\theta)/2$, $b_1=b_3=\theta$,
$b_2=(1-2\theta)$, $b_4=0$, and $\theta=1/(2-\sqrt[3]{2}) \approx 1.3512$.
Propagation (8) can be related to the position version of the FR integrator
(PFR), because putting formally $\theta = 0$ it transforms to the
second-order PV algorithm (7). For $A \leftrightarrow B$, Eq.~(8) will
represent the velocity FR counterpart (VFR), which can be derived directly
from Eq.~(5) at $a_1=0$, $a_2=a_4=\theta$, $a_3=(1-2\theta)$, $b_1=b_4=
\theta/2$ and $b_2=b_3=(1-\theta)/2$.

\section{Optimization of fourth-order algorithms within direct
         decomposition}

Let us consider now an extended decomposition scheme of the fourth order
by allowing to accept a value for $P$ which exceeds the necessary minimum
($P=4$) on unity, i.e., letting $P=5$. Remember that we cannot choose the
number $P$ to be too big, because this results in too larger number, namely
$P-1$, of expensive force recalculations. Chossing $P=5$ we hope simply
to reduce the truncation errors ${\cal O}(h^5)$ significantly in a little
additional computation cost, rather than to increase the order of the
decomposition scheme (note that sixth-order integrators are derivable
\cite{Yoshida} beginning up from $P=8$).

For $P=5$, the extended decomposition can be presented in the form
\begin{eqnarray}
{\rm e}^{(A+B) h + C_3 h^3 + C_5 h^5 + {\cal O}(h^7)} =
{\rm e}^{B \xi h}
{\rm e}^{A (1 - 2 \lambda) \frac{h}{2}}
{\rm e}^{B \chi h}
&& \times \nonumber \\ [-6pt] \\ [-6pt]
{\rm e}^{A \lambda h}
{\rm e}^{B (1 - 2 (\chi + \xi)) h}
{\rm e}^{A \lambda h}
{\rm e}^{B \chi h}
{\rm e}^{A (1 - 2 \lambda) \frac{h}{2}}
{\rm e}^{B \xi h} \nonumber && \ \
\end{eqnarray}
following from Eq.~(5) at $a_1=0$, $b_1=b_5=\xi$, $a_2=a_5=(1-2\lambda)/2$,
$b_2=b_4=\chi$, $a_3=a_4=\lambda$, and $b_3=1-2 (\chi+\xi)$. Here the
symmetry of time coefficients and the condition $\sum_{p=1}^5 a_p=
\sum_{p=1}^5 b_p=1$ have already been taken into account. Again, the
propagation with $A \leftrightarrow B$ is also acceptable, and then $a_1=
a_5=\xi$, $b_1=b_5=(1-2\lambda)/2$, $a_2=a_4=\chi$, $b_3=b_4=\lambda$,
$a_3=1-2 (\chi+\xi)$, and $b_5=0$. The operator $C_3$, appearing in
the left-hand-side of Eq.~(9), is responsible for the cancellation of
third-order truncation uncertainties. The explicit expression for it is
\begin{equation}
C_3 = \alpha(\xi,\lambda,\chi) [A,[A,B]] +
\beta(\xi,\lambda,\chi) [B,[A,B]] \, ,
\end{equation}
where $[ \ , \ ]$ denotes the commutator of two operators, and
\begin{eqnarray}
\alpha(\xi,\lambda,\chi)&=&-\frac{1}{24}+\lambda^2 \chi+\frac{\xi}{4} \, ,
\nonumber \\ [-6pt] \\ [-6pt]
\beta(\xi,\lambda,\chi)&=&-\frac{1}{12}+\lambda \chi (1-\chi-2 \xi)+
\frac{\xi}{2} - \frac{\xi^2}{2} \, . \nonumber
\end{eqnarray}
So that at $C_3=0$, i.e. when $\alpha(\xi,\lambda,\chi)=0$ and $\beta(\xi,
\lambda,\chi)=0$, formula (9) represents a whole family of symplectic
time-reversible integrators of the fourth-order.

A particular member of the above family can be obtained by choosing
corresponding values for $\xi$, $\lambda$, and $\chi$. As far as there
are three parameters and only two constraints, $\alpha=0$ and $\beta=0$,
one from these parameters, $\xi$ say, can be treated to be free. Then,
for example, putting $\xi=0$, Eq.~(9) reduces to the original FR algorithm
(8) in position or (when $A \leftrightarrow B$) velocity forms. The extended
(when $\xi \ne 0$) propagation will require already four, instead of three,
force recalculation per time step. However, having a room in varying $\xi$,
we can overcompensate the increased computational efforts by minimizing
the fifth-order truncation uncertainties $C_5 h^5$.

In order to show this, let us analyze in detail the influence of these
uncertainties on the result. Expanding both the sides of Eq.~(9) into
Taylor's series with respect to $h$, one finds
\begin{eqnarray} \label{coef}
C_5 = \gamma_1 [A,[A,[A,[A,B]]]] &\!+\!&
      \gamma_2 [A,[A,[B,[A,B]]]] \!+\!
      \nonumber \\
      \gamma_3 [B,[A,[A,[A,B]]]] &\!+\!&
      \gamma_4 [B,[B,[B,[A,B]]]] \!+\!
                \\
      \gamma_5 [B,[B,[A,[A,B]]]] &\!+\!&
      \gamma_6 [A,[B,[B,[A,B]]]] \, ,
      \nonumber
\end{eqnarray}
where explicit expressions for $\gamma$-multipliers are:
\begin{eqnarray}
\gamma_1&=&\frac{7}{5760} - \frac{\lambda^2 \chi}{12} \bigg(
\frac{1}{2} - \lambda^2 \bigg) - \frac{\xi}{192} \, , \nonumber \\
\gamma_2&=&\frac{1}{480} - \frac{\lambda \chi}{2} \bigg(
\frac{1}{12} - \frac{\lambda}{6} + \lambda^2 \chi + \lambda \xi -
\frac{\chi}{12} - \frac{\xi}{6} \bigg) - \frac{\xi^2}{24} \, , \nonumber \\
\gamma_3&=&\frac{1}{360} - \lambda^2 \chi \bigg(
\frac{1}{6} - \frac{\lambda}{6} - \frac{\lambda \chi}{3} +
\frac{\lambda \xi}{3} - \frac{\xi}{2} \bigg) -
\frac{\xi}{48} + \frac{\xi^2}{24} \, , \nonumber \\
\gamma_4&=&\frac{1}{720} - \lambda \chi \bigg(
\frac{\chi}{12} - \frac{\chi^2}{6} + \frac{\chi^3}{12} -
\frac{\chi \xi}{2} + \frac{\chi^2 \xi}{3} + \frac{\chi \xi^2}{2} +
\nonumber \\
&& \ \ \ \ \ \ \ \ \ \ \ \ \ \ \ \,
\frac{\xi}{6} - \frac{\xi^2}{2} + \frac{\xi^3}{3} \bigg) -
\frac{\xi^2}{24} + \frac{\xi^3}{12} - \frac{\xi^4}{24} \, , \nonumber \\
\gamma_5&=&\frac{1}{120} - \lambda \chi \bigg(
\frac{1}{6} - \frac{\lambda}{2} \bigg[ \frac{1}{6} + \frac{\chi}{2} -
\chi^2 - \chi \xi - \xi + \xi^2 \bigg] -
\nonumber \\
&& \ \ \ \ \ \ \ \ \ \ \ \ \ \ \
\frac{\chi}{6} +
\frac{\chi \xi}{2} - \frac{5 \xi}{6} + \xi^2 \bigg) -
\frac{\xi}{16} + \frac{7 \xi^2}{48} - \frac{\xi^3}{8} \, , \nonumber \\
\gamma_6&=&-\frac{1}{360} + \lambda \chi \bigg(
\frac{1}{12} - \frac{\lambda \chi}{2} + \frac{2 \lambda \chi^2}{3} +
\lambda \chi \xi - \frac{\chi}{12} +
\nonumber \\
&& \ \ \ \ \ \ \ \ \ \ \ \ \ \ \ \ \
\frac{\chi \xi}{2} -
\frac{2 \xi}{3} + \xi^2 \bigg) + \frac{\xi}{24} -
\frac{\xi^2}{6} + \frac{\xi^3}{6} \, . \nonumber
\end{eqnarray}
Assuming that all the fifth-order commutators in (\ref{coef}) are
nonzero valued, the norm of $C_5$ with respect to fifth-order
commutators arising in Eq.~(12) can be written as follows
\begin{equation}
\gamma(\xi,\lambda,\chi)=\sqrt{\gamma_1^2+\gamma_2^2+\gamma_3^2+
\gamma_4^2+\gamma_5^2+\gamma_6^2} \, .
\end{equation}
Then the norm of local uncertainties $C_5 {\mbox{\boldmath $\rho$}} h^5$
appearing in phase trajectory ${\mbox {\boldmath $\rho$}}$ during a
single-step propagation given by Eqs.~(4) and (9) can be expressed in
terms of $\gamma$ and $h$ as $g=\gamma h^5$. During a whole integration
over a fixed time interval $t$, the total number $l$ of such single steps
is proportional to $h^{-1}$. As a result, the local fifth-order
uncertainties will accumulate step by step leading at $t \gg h$ to the
fourth-order global errors $\Gamma=g h^{-1}$, i.e.,
\begin{equation}
\Gamma(h,\xi,\lambda,\chi) = \gamma(\xi,\lambda,\chi) h^4 \, .
\end{equation}

Extended propagation (9) can now be optimized with respect to time
coefficients $\xi$, $\lambda$, and $\chi$ by finding the global minimum
for the function $\gamma(\xi,\lambda,\chi)$, provided $\alpha=0$ and
$\beta=0$, i.e., solving the system of equations
\begin{equation}
\left \{
\begin{array}{l}
\alpha(\xi,\lambda,\chi)=0 \, , \\ [3pt]
\beta(\xi,\lambda,\chi)=0 \, ,  \\  [3pt]
\gamma(\xi,\lambda,\chi)=\min {\rm (global)} \, .
\end{array}
\right.
\end{equation}
A way to simplify the problem is to solve analytically the first two
equations of (15) with respect to $\chi$ and $\xi$,
\begin{eqnarray}
\chi(\lambda)&=&\frac{4 \lambda - 8 \lambda^2 \pm
\sqrt{2 \lambda (-1 + 8 \lambda - 24 \lambda^2 + 24 \lambda^3)}}
{12 \lambda - 96 \lambda^3 + 96 \lambda^4} \, , \ \ \ \ \
\nonumber \\ [-6pt] \\ [-6pt]
\xi(\lambda)&=&\frac16 - 4 \lambda^2 \chi \, , \nonumber
\end{eqnarray}
then substitute expressions (16) into the third equation, and find
numerically the minimum considering already $\gamma(\xi(\lambda),
\lambda,\chi(\lambda))$ as a function of only one variable $\lambda$.
The result within sixteenth significant digits is
\begin{eqnarray}
\xi    &=& +0.1720865590295143{\rm E}\!+\!00 \nonumber \\
\lambda&=& -0.9156203075515678{\rm E}\!-\!01           \\
\chi   &=& -0.1616217622107222{\rm E}\!+\!00 \, . \nonumber
\end{eqnarray}

The global minimum of $\gamma(\xi,\lambda,\chi)$, corresponding to
solutions (17) found (they satisfy Eq.~(16) at sign ``$+$'' in the
right hand side of expression for $\chi(\lambda)$), consists
$\gamma_{\rm min}^{\rm EFRL} \approx 0.00092$, where the
superscript EFRL refers to the extended FR-like integration (9).
On the other hand, the value of $\gamma$ corresponding to usual FR
scheme (8), i.e. when $\xi=0$, $\lambda=(1-\theta)/2$ and $\chi=
\theta$ (then Eq.~(9) reduces to Eq.~(8)), is equal to
$\gamma_{\rm FR} \approx$ 0.039. We see, therefore, that applying
the extended integration allows one to decrease the truncation
errors approximately in $\gamma_{\rm FR}/\gamma_{\rm min}^{\rm
EFRL} \approx 42$ times. Taking into account that such a
decreasing has been achieved increasing the number of force
evaluations per step from three to four, the extended propagation
must be performed with step sizes which are in factor 4/3 higher
than those of the FR algorithm, in order to provide the same
number of total force recalculations during the fixed overall
interval of integration. Thus, we will come to more efficient
calculations if the following inequality $\Gamma_{\rm min}^{\rm
EFRL}(4h/3) < \Gamma_{\rm FR}(h)$ takes place. In view of
Eq.~(14), such an inequality can be rewritten as $\gamma_{\rm FR}/
\gamma_{\rm min}^{\rm EFRL} > (4/3)^4 \approx$ 3.16, so that it is
fulfilled completely in the optimization regime. In particular,
\begin{equation}
\frac{\Gamma_{\rm min}^{\rm EFRL}(4h/3)}{\Gamma_{\rm FR}(h)} \approx 0.075
\end{equation}
indicating that the global errors can be reduced more than in 10 times
with respect to the FR integration without spending any additional
overall computational costs.

The proposed procedure can be used, in principle, for the decompositions
of arbitrary two noncommutative operators $A$ and $B$ to achieve the best
performance. Note that the necessity in these decompositions may arise not
only when considering the integration of motion in classical systems, but
also in mathematical and quantum mechanical calculations. In some cases,
further improvement of the efficiency of the computations may also be
possible. For instance, an extra optimization of the decomposition scheme
can be achieved in celestial mechanics applications due to a specific
character of motion in the solar system (see Appendix).

In the case of MD simulations with velocity-indepen\-dent forces, an
additional optimization can be carried out as well using specific
properties of operators $A$ and $B$. Taking into account explicit
expressions for these operators, it can be verified readily that
two of the sixth five-order commutators vanish in Eq.~(12), namely,
$[B,[B,[B,[A,B]]]]=0$ and $[A,[B,[B,[A,B]]]]=0$, when time propagation
is performed with the help of extended scheme (9). This scheme, by
analogy to Verlet integrators, we will refer to the velocity version of
the EFRL algorithm and abbreviate in our notations as VEFRL. Then,
letting formally $\gamma_4=0$ and $\gamma_6=0$ to exclude the above
zeroth commutators in Eq.~(13), and resolving problem (15) on condition
minimum yields
\begin{eqnarray}
\xi    &=& +0.1644986515575760{\rm E}\!+\!00 \nonumber \\
\lambda&=& -0.2094333910398989{\rm E}\!-\!01           \\
\chi   &=& +0.1235692651138917{\rm E}\!+\!01 \, . \nonumber
\end{eqnarray}
This leads to the global minimum $\gamma_{\rm min}^{\rm VEFRL} \approx
0.00065$ (which is achieved at sign ``$-$'' in Eq.~(16)), while the
value of $\gamma$ corresponding to usual FR integration (8) reduces (at
$\gamma_4=0$ and $\gamma_6=0$) to $\gamma_{\rm FR} \approx$ 0.028. Again,
the truncation uncertainties decrease approximately in the same factor,
$\gamma_{\rm FR}/\gamma_{\rm min}^{\rm VEFRL} \approx 43$, and
$\Gamma_{\rm min}^{\rm VEFRL}(4h/3)/\Gamma_{\rm FR}(h) \approx 0.073$.

The position EFRL integration (PEFRL) is obtained from Eq.~(9) by
replacing $A \leftrightarrow B$. In such a case, the zeroth five-order
commutators are $[A,[A,[A,[A,B]]]]=0$ and $[B,[A,[A,[A,B]]]]$. They can
be excluded in Eq.~(13) by putting $\gamma_1=0$ and $\gamma_3=0$. Then
solutions to system (15) transform into
\begin{eqnarray}
\xi    &=& +0.1786178958448091{\rm E}\!+\!00 \nonumber \\
\lambda&=& -0.2123418310626054{\rm E}\!+\!00           \\
\chi   &=& -0.6626458266981849{\rm E}\!-\!01 \nonumber
\end{eqnarray}
(with sign ``$+$'' in Eq.~(16)), and the minimum is $\gamma_{\rm min}^{\rm
PEFRL} \approx 0.00061$. The value of $\gamma$ corresponding to original
scheme (8) is now (when $\gamma_1=0$ and $\gamma_3=0$) equal to $\gamma_{\rm
FR} \approx$ 0.038, so that $\gamma_{\rm FR}/\gamma_{\rm min}^{\rm PEFRL}
\approx 62$ and $\Gamma_{\rm min}^{\rm PEFRL}(4h/3)/\Gamma_{\rm FR}(h)
\approx 0.051$.

In view of Eqs.~(4), (6), and (9), more explicit expressions for the
single-step propagation of position and velocity from time $t$ to $t+h$
within the optimized VEFRL algorithm are:
\begin{eqnarray}
{\bf v}_1&=&{\bf v}(t)+{\textstyle\frac1m}{\bf f}[{\bf r}(t)] \xi h
\nonumber \\ [1pt]
{\bf r}_1&=&{\bf r}(t)+{\bf v}_1 (1 - 2 \lambda) h/2
\nonumber \\ [1pt]
{\bf v}_2&=&{\bf v}_1+{\textstyle\frac1m}{\bf f}[{\bf r}_1] \chi h
\nonumber \\ [1pt]
{\bf r}_2&=&{\bf r}_1+{\bf v}_2 \lambda h
\nonumber \\ [1pt]
{\bf v}_3&=&{\bf v}_2+{\textstyle\frac1m}{\bf f}[{\bf r}_2]
(1 - 2 (\chi + \xi)) h
\\ [1pt]
{\bf r}_3&=&{\bf r}_2+{\bf v}_3 \lambda h
\nonumber \\ [1pt]
{\bf v}_4&=&{\bf v}_3+{\textstyle\frac1m}{\bf f}[{\bf r}_3] \chi h
\nonumber \\ [1pt]
{\bf r}(t+h)&=&{\bf r}_3+{\bf v}_4 (1 - 2 \lambda) h/2
\nonumber \\ [1pt]
{\bf v}(t+h)&=&{\bf v}_4+{\textstyle\frac1m}{\bf f}[{\bf r}(t+h)] \xi h
\nonumber
\end{eqnarray}
where the values for $\xi$, $\lambda$, and $\chi$ should be taken from
Eq.~(19). The optimized PEFRL algorithm (when $A \leftrightarrow B$ in
Eq.~(9)) reads:
\begin{eqnarray}
{\bf r}_1&=&{\bf r}(t)+{\bf v}(t) \xi h
\nonumber \\ [1pt]
{\bf v}_1&=&{\bf v}(t)+{\textstyle\frac1m}{\bf f}[{\bf r}_1]
(1 - 2 \lambda) h/2
\nonumber \\ [1pt]
{\bf r}_2&=&{\bf r}_1+{\bf v}_1 \chi h
\nonumber \\ [1pt]
{\bf v}_2&=&{\bf v}_1+{\textstyle\frac1m}{\bf f}[{\bf r}_2]\lambda h
\nonumber \\ [1pt]
{\bf r}_3&=&{\bf r}_2+{\bf v}_2 (1 - 2 (\chi + \xi)) h
\\ [1pt]
{\bf v}_3&=&{\bf v}_2+{\textstyle\frac1m}{\bf f}[{\bf r}_3]\lambda h
\nonumber \\ [1pt]
{\bf r}_4&=&{\bf r}_3+{\bf v}_3 \chi h
\nonumber \\ [1pt]
{\bf v}(t+h)&=&{\bf v}_3+{\textstyle\frac1m}{\bf f}[{\bf r}_4]
(1 - 2 \lambda) h/2
\nonumber \\ [1pt]
{\bf r}(t+h)&=&{\bf r}_4+{\bf v}(t+h)] \xi h
\nonumber
\end{eqnarray}
and the parameters $\xi$, $\lambda$, and $\chi$ should accept their
values from Eq.~(20). The algorithms are explicit, simple in implementation,
and require only slight modification with respect to the original FR scheme.

\section{Optimization by composing second-order schemes}

Another way to construct high-order algorithms consists in composing
lower-order schemes. In particular, employing second-order Verlet integrator
(7), the composition can be performed \cite{Suzukium} for an arbitrary
higher order $K > 2$ as
\begin{equation}
{\rm e}^{(A+B)h + {\cal O}(h^{K+1})} = \prod_{q=1}^{Q}
{\rm e}^{A d_q \frac{h}{2}} {\rm e}^{B d_q h}
{\rm e}^{A d_q \frac{h}{2}} \equiv \prod_{q=1}^{Q} S_2(d_q h) \, .
\end{equation}
Here the coefficients $d_q$ are chosen by providing a maximum for $K$
at a given number $Q>1$. This leads to the necessity of fulfilling a
set of order constraints, and, for instance, the condition $\sum_{q=1}^Q
d_q=1$ must be satisfied to avoid the first-order truncation terms in
${\cal O}(h^{K+1})$. For time-reversible compositions, the coefficients
$d_q$ should appear symmetrically, i.e., satisfy the property $d_q=
d_{Q-q+1}$. Then, as in the case of direct decomposition (5), even-order
terms will be absent in the function ${\cal O}(h^{K+1})$, and the order
of reversible composition schemes will also accept only even numbers.

Fourth-order ($K=4$) composition integrators are derivable from Eq.~(23)
at $Q \ge 3$. For example, putting $Q=3$ as well as $d_1=d_3=\theta$ and
$d_2=1-2\theta$ one comes to the scheme
\begin{equation}
{\rm e}^{(A+B) h + {\cal O}(h^5)} = S_2(\theta h)
S_2\big((1-2\theta) h\big) S_2(\theta h)
\end{equation}
which coincides entirely with the FR algorithm (as can be seen by comparing
with Eq.~(8)). Thus, in this particular situation, the direct decomposition
and second-order-based composition approaches exhibit to be identical (but
with increasing $K$ or $Q$ both the approaches will lead to different
results). We will now consider the construction of extended (when $Q > 3$)
composition algorithms of the fourth order.

It can be shown that no real solutions exist for $d_q$ at $Q=4$ and $K=4$.
So putting $Q=5$, one obtains the simplest extended fourth-order
composition formula
\begin{eqnarray}
{\rm e}^{(A+B) h + C_3 h^3 + C_5 h^5 + {\cal O}(h^7)} =
\ \ \ \ \ \ \ \ \ \ \ \ \ \ \ \ \ \ \ \ \ \ \ \ \ \
\nonumber \\ [-6pt] \\ [-6pt] \ \ \ \
S_2(\xi h) S_2(\lambda h)
S_2((1-2(\xi+\lambda) h) S_2(\lambda h) S_2(\xi h) \, ,
\nonumber
\end{eqnarray}
following from Eq.~(23) at $d_1=d_5=\xi$, $d_2=d_4=\lambda$, and
$d_3=1-2(\xi+\lambda)$, where
\begin{equation}
C_3 = - \alpha(\xi,\lambda) \Big( \frac{1}{24}
[A,[A,B]] + \frac{1}{12} [B,[A,B]] \Big)
\end{equation}
with
\begin{equation}
\alpha(\xi,\lambda) = 2 \xi^3 + 2 \lambda^3 +
\big( 1 - 2 (\xi+\lambda) \big)^3 \, .
\end{equation}
Formula (25) constitutes a family of composition time-reversible integrators
of the fourth-order, provided $C_3=0$, i.e. $\alpha(\xi,\lambda)=0$. Any one
from the two parameters $\xi$ and $\lambda$ can be chosen, in principle,
arbitrarily since we have only one constraint. For $\xi=0$ and $\lambda=
\theta$, we come to usual FR integration (8). Suzuki \cite{Suzukium},
considering an extended scheme like (25), has imposed the additional
constraint $\xi=\lambda \equiv \vartheta$, and obtained $\vartheta=1/
(4-\sqrt[3]{4}) \approx 0.41449$. We will show below that although
Suzuki's approach leads to an increased efficiency with respect to the
original FR scheme, it is not the best choice for fourth-order integration.

The operator $C_5$, which forms the fifth-order term of truncation
uncertainties in composition (25), can again be presented in form of
Eq.~(12), where now
\begin{eqnarray}
\gamma_1 &=&\frac{7 s}{5760} - \frac{w}{24}   , \ \
\gamma_2  =\frac{s}{480} - \frac{w}{8}       , \ \
\gamma_3  =\frac{s}{360} - \frac{w}{24}      , \ \ \ \ \
\nonumber \\ [0pt] \\ [-3pt]
\gamma_4 &=& \frac{s}{720} - \frac{w}{12}     , \ \
\gamma_5  = \frac{s}{120} - \frac{w}{8}      , \ \
\gamma_6  =-\frac{s}{360} - \frac{w}{12}
\nonumber
\end{eqnarray}
with
\begin{eqnarray*}
s&(&\xi,\lambda) =
       1-10 \lambda (1-8 \xi+24 \xi^2-32 \xi^3+16 \xi^4)+
                   \\
      && 40 \lambda^2 (1-6 \xi+12 \xi^2-8 \xi^3)-
         80 \lambda^3 (1-4 \xi+4 \xi^2)+
                   \\
      && 80 \lambda^4 (1-2 \xi)-30 \lambda^5-
         10 \xi (1-4 \xi+8 \xi^2-8 \xi^3+3 \xi^4)
\end{eqnarray*}
and
\begin{eqnarray*}
w&(&\xi,\lambda) =
          (-\lambda (1-14 \xi+54 \xi^2-80 \xi^3+40 \xi^4)+
                   \\
        &&  \lambda^2 (7-54 \xi+120 \xi^2-80 \xi^3)-
            \lambda^3 (17-74 \xi+74 \xi^2)+
                   \\
        &&  \lambda^4 (17-34 \xi)-6 \lambda^5-
            \xi+7 \xi^2-17 \xi^3+17 \xi^4-6 \xi^5)/6 \, .
\end{eqnarray*}
It is worth remarking that the operators $C_3$ and $C_5$ cannot be reduced
to zero simultaneously at $Q=5$, since we cannot satisfy all the three
equations $\alpha=0$, $s=0$, and $w=0$ having only two parameters $\xi$
and $\lambda$ (such a reduction, resulting in sixth-order composition
integrators, is possible within real coefficients $d_q$ beginning up
from $Q=7$ \cite{Suzukium}).

Our task is to minimize the norm
\begin{equation}
\gamma(\xi,\lambda)=\sqrt{\gamma_1^2+\gamma_2^2+\gamma_3^2+
\gamma_4^2+\gamma_5^2+\gamma_6^2}
\end{equation}
of $C_5$ in space of fifth-order commutators, provided $C_3=0$, i.e.,
to solve the problem
\begin{equation}
\left \{
\begin{array}{l}
\alpha(\xi,\lambda)=0 \, , \\ [3pt]
\gamma(\xi,\lambda)=\min {\rm (global)} \, .
\end{array}
\right.
\end{equation}
The simplest way to do this is to transform the first of two equalities
of system (30) from the cubic (see Eq.~(27)) to square equation replacing
the variable $\xi$ by the new independent quantity $\chi=1-2(\xi+\lambda)$,
and find solutions for $\lambda$ considering $\chi$ as a parameter. As a
result, we obtain
\begin{eqnarray}
\lambda(\chi)&=& \frac{ 3 (1 - \chi)^2 \!\pm\!
\sqrt{3 (-1 + 4 \chi - 6 \chi^2 - 12 \chi^3 + 15 \chi^4)}}
     {12 (1 - \chi)} \, , \nonumber
     \\ [-2pt] \\ [-2pt]
\xi(\chi)&=&\big( 1-2 \lambda(\chi)-\chi \big)/2 \, .
\nonumber
\end{eqnarray}
Then, substituting these expressions into the second equation, we come to
the function $\gamma\big(\xi(\chi),\lambda(\chi)\big) \equiv \gamma(\chi)$
which depends already on only one variable $\chi$. The global minimum of
this function is achieved at $\chi=-0.7269082885036828{\rm E}\!+\!00$ (with
sign ``$+$'' for $\lambda(\chi)$ in Eq.~(31)) and consists $\gamma_{\rm
min}^{\rm ESL} \approx 0.0011$, where the superscript ESL refers to the
extended Suzuki-like integration (25). So that according to Eq.~(31), the
corresponding solutions are
\begin{eqnarray}
\xi    &=& 0.3221375960817984{\rm E}\!+\!00
\nonumber \\ [-7pt] \\ [-7pt]
\lambda&=& 0.5413165481700430{\rm E}\!+\!00 \, .
\nonumber
\end{eqnarray}

As was mentioned above, Suzuki has used the parameters $\xi=\lambda=
\vartheta \approx 0.41449$ in his integration. This corresponds to the
value $\gamma_{\rm S} \approx 0.0015$ of function (29) that is approximately
in factor 1.5 higher than its optimized counterpart $\gamma_{\rm min}^{\rm
ESL} \approx 0.0011$. Therefore, his choice was not optimal. But the main
conclusion we may make in this context is that the original Suzuki approach
as well its optimized ESL version are less efficient than the EFRL
integration described in the preceding section. Indeed, taking into account
the value $\gamma_{\rm min}^{\rm EFRL} \approx 0.00092$ and the fact that
the Suzuki approach requires up five (see Eq.~(25)), instead of four, force
evaluations per time step, we obtain in view of Eq.~(14) that efficiency of
the EFRL integration is in factor
\begin{equation}
\frac{\Gamma_{\rm min}^{\rm ESL}(5h/4)}
{\Gamma_{\rm min}^{\rm EFRL}(h)} =
\frac{\gamma_{\rm min}^{\rm ESL}}{\gamma_{\rm min}^{\rm EFRL}}
\left( \frac{5}{4} \right)^4 \approx 3
\end{equation}
better with respect to the ESL scheme, and, thus, approximately in 5 times
higher with respect to the original approach by Suzuki.

Note that results (32) and (33) correspond to a general case of integration
of motion when forces may explicitly depend, rigorously speaking, on
velocities. For velocity-independent accelerations, we can exploit
additional properties of operators $A$ and $B$, to improve the efficiency
of the compositions. Mention that such properties are $[B,[B,[B,[A,B]]]]=0$
and $[A,[B,[B,[A,B]]]]$, or $[A,[A,[A,[A,B]]]]=0$ and $[B,[A,[A,[A,B]]]]$
at $A \leftrightarrow B$, and they can be taken into account by putting
formally $\gamma_4=0$ and $\gamma_6=0$, or $\gamma_1=0$ and $\gamma_3=0$
in Eq.~(29). Then within our basic definitions for $A$ and $B$, Eq.~(25)
will lead to the position version of the ESL integration (PESL). For this
version the optimal solutions we found are
\begin{eqnarray}
\xi    &=& 0.3162227486360109{\rm E}\!+\!00
\nonumber \\ [-7pt] \\ [-7pt]
\lambda&=& 0.5521563637246984{\rm E}\!+\!00
\nonumber
\end{eqnarray}
with the minimum value $\gamma_{\rm min}^{\rm PESL} \approx 0.00109$ that
corresponds to $\chi=-0.7367582247214187{\rm E}\!+\!00$ (and sign ``$+$''
in Eq.~(31)). Remembering that for the position version of the EFRL
integration was $\gamma_{\rm min}^{\rm PEFRL} \approx 0.00061$, one
obtains $\Gamma_{\rm min}^{\rm PESL}(5h/4)/\Gamma_{\rm min}^{\rm PEFRL}(h)
\approx 4$, whereas with respect to the position version of original
Suzuki's approach (PS) for which $\gamma_{\rm PS} \approx 0.0014$, the
efficiency increases more than in 5 times. It is necessary to point out
that solutions (34) correspond to a local minimum of $\gamma$ (when
$\gamma_4=0$ and $\gamma_6=0$). The global minimum is achieved at
$\xi =  1.453335388755048$, $\lambda = -2.154220603244978$, with $\chi=
2.401770428979862$ and consists about 0.00096 that is only slightly smaller
than the presented above value 0.00109. But now all the time coefficients
appear to be significantly larger than their counterparts (34). In such a
situation, the preference should be given to the local minimum, because it
leads to a significantly smaller value for the norm of higher-order
truncation uncertainties ${\cal O}(h^7)$, and thus the precision of
composition (25) will be much less sensitive to increasing the size of
the time step $h$ (see the next section). For $A \leftrightarrow B$,
Eq.~(25) will correspond to the velocity version of the ESL integration
(VESL) and its solutions (at $\gamma_1=0$ and $\gamma_3=0$) to Eq.~(30)
take the values
\begin{eqnarray}
\xi    &=& 0.3226106225667342{\rm E}\!+\!00
\nonumber \\ [-7pt] \\ [-7pt]
\lambda&=& 0.5404642725582767{\rm E}\!+\!00
\nonumber
\end{eqnarray}
with $\chi=-0.7261497902500218{\rm E}\!+\!00$ (sign ``$+$'' in Eq.~(31))
and the global minimum $\gamma_{\rm min}^{\rm VESL} \approx 0.00107$. At
the same time, $\gamma_{\rm min}^{\rm VEFRL} \approx 0.00065$ and for the
velocity version of Suzuki's approach (VS) we obtain $\gamma_{\rm VS}
\approx 0.0014$. So that $\Gamma_{\rm min}^{\rm VESL}(5h/4)/\Gamma_{\rm
min}^{\rm VEFRL}(h) \approx 4$, and again the ratio of efficiencies of
the VEFRL and original VS scheme is equal approximately to 5.

\section{Application to molecular dynamics simulations}

In order to verify theoretical predictions presented in sections III
and IV, we have applied the velocity and position versions of our EFRL
and ESL algorithms to MD simulations of a Lennard-Jones fluid (LJ),
and compared the results with those of the original FR and Suzuki's
approaches. The system under consideration was a collection of $N=256$
particles interacting through a shifted LJ-like potential, $\varphi(r)=
\Phi(r)-\Phi(r_{\rm c})$ at $r < r_{\rm c}$ and $\varphi(r)=0$ otherwise,
where $\Phi(r)= 4 u \big[(\sigma/r)^{12}-(\sigma/r)^6\big]$ is the genuine
LJ potential. The particles were placed in a basic cubic box of volume
$V=L^3$, and the modification of $\Phi(r)$ with $r_{\rm c}=L/2 \approx
3.36 \sigma$ as well as the periodic boundary conditions have been used
to exclude the finite-size effects. The simulations were performed in
a microcanonical ensemble at a reduced density of $n^\ast=\frac{N}{V}
\sigma^3=0.845$ and a reduced temperature of $T^\ast=k_{\rm B}T/u=1.7$.
All runs of the length in $l=10\,000$ time steps each were started from
an identical well equilibrated initial configuration ${\mbox{\boldmath
$\rho$}}(0)$. The precision of the algorithms was measured in terms of
the relative total energy fluctuations ${\cal E}=\langle (E- \langle E
\rangle)^2 \rangle/|\langle E \rangle|$, where $E=\frac12 \sum_{i=1}^N
m {{\bf v}_i}^2/2+\frac12 \sum_{j (j \ne i)}^N \varphi(r_{ij})$ and
$\langle \ \rangle$ denotes the microcanonical averaging. Note that if
the equations of motion could be solved exactly, the above fluctuations
should vanish, because in microcanonical ensembles the total energy is
an integral of motion, $E(t) = E(0)$. So that during approximate MD
integrations, smaller values of ${\cal E}$ will correspond to a better
precision in evaluation of phase trajectories.

\begin{figure}[htbp]
\begin{centering}
\begin{picture}(84,103)
\epsfxsize=84mm \put(0,0){ \epsffile[41 501 561 702]{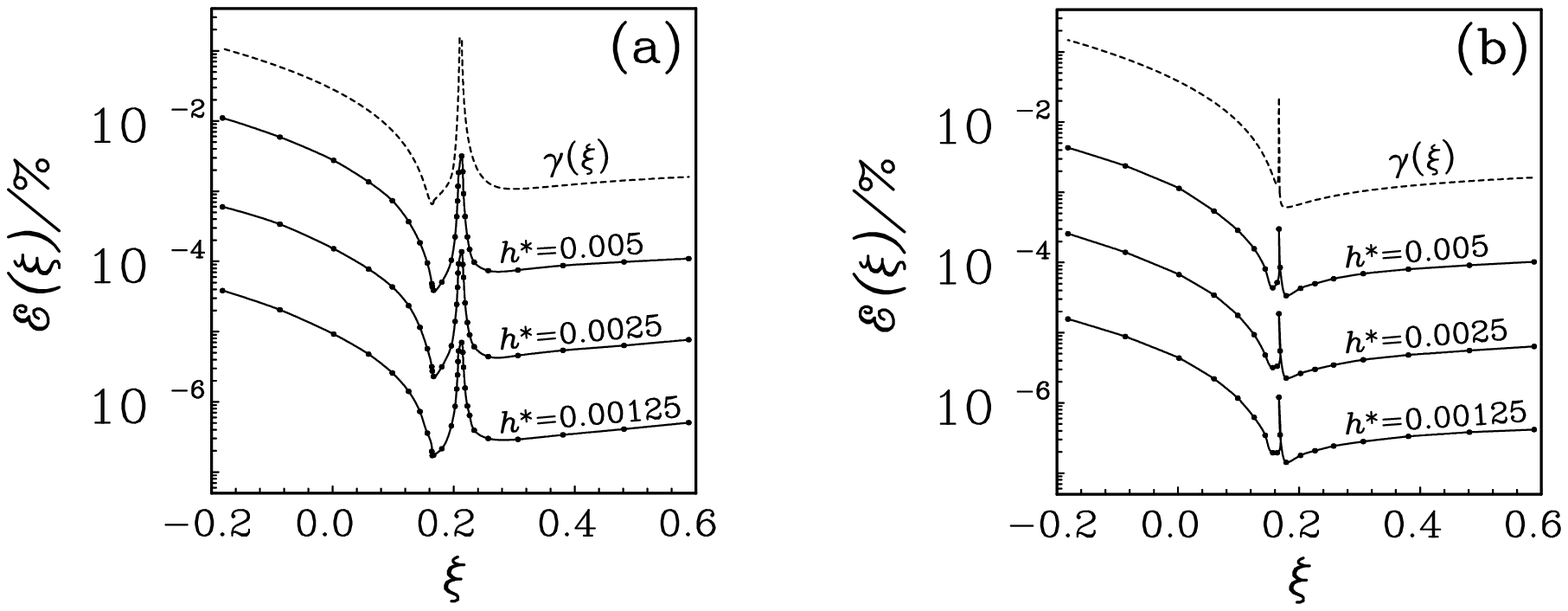}}
\end{picture}
\end{centering}
\end{figure}
{\small FIG. 1. The total energy fluctuations obtained in the
simulations for different values of free parameter $\xi$ at three
fixed time steps, $h^\ast=0.00125$, 0.0025, and 0.005 using the
VEFRL (subset (a)) and PEFRL (subset (b)) integration (Eqs.~(21)
and (22, respectively). The simulation results are presented by
circles connected by the solid curves. The function $\gamma(\xi)$
(see Eq.~(13)) is plotted by the dashed curves.}

\vspace{7pt}

The equations of motion were first integrated with the help of extended
decomposition scheme (9) in which the parameters $\xi$, $\lambda$, and
$\chi$, being constant within each run, varied from one run to another.
For convenience of presentation, we have chosen $\xi$ (instead of
$\lambda$ as earlier) to be a free parameter, so that the two other
quantities $\lambda(\xi)$ and $\chi(\xi)$ should be treated as depending
on $\xi$ according to constraints (16). The total energy fluctuations
obtained in such simulations at the end of the runs for three (fixed within
each run) undimensional time steps, namely $h^\ast=h (u/m\sigma^2)^{1/2}=
0.00125$, 0.0025, and 0.005, are shown in Fig.~1 as functions of parameter
$\xi$. The subsets (a) and (b) of this figure correspond to VEFRL (Eq.~(21))
and PEFRL (Eq.~(22)) versions, respectively. As can be seen, all the
dependencies ${\cal E}(\xi,h)$ have the global minimum which locates at
the same point, $\xi \approx 0.164$ for VEFRL or $\xi \approx 0.179$ for
PEFRL, independently on the size $h$ of the time step. This point coincides
completely with the minimum given by Eq.~(19) or (20) for the function
$\gamma(\xi) \equiv \gamma(\xi,\lambda(\xi),\chi(\xi))$ (see Eq.~(13), in
which we should put $\gamma_4=0$ and $\gamma_6=0$ for VEFRL or $\gamma_1=0$
and $\gamma_3=0$ for PEFRL) which is also included in Fig.~1 (dashed curves
in the subsets). Moreover, the fluctuations ${\cal E}(\xi,h)$ appear to be
proportional to the norm $\Gamma=\gamma h^4$ (Eq.~(14)) of global errors,
and the coefficient of this proportionality almost does not depend on $\xi$
and $h$. In addition, at each time step the energy fluctuations decrease
at the minimum in 40--60 times (in dependence on the versions considered)
with respect to those at $\xi=0$, that is in agreement with our predicted
values $\gamma_{\rm FR}/\gamma_{\rm min}^{\rm VEFRL} \approx 43$ and
$\gamma_{\rm FR}/\gamma_{\rm min}^{\rm PEFRL} \approx 62$.

The result for the total energy fluctuations as functions of the
length $l=t/h$ of the simulations corresponding to the optimized
VEFRL (Eqs.~(19) and (21))) and PEFRL (Eqs.~(20) and (22))
algorithms is presented in subsets (a), (b), (c), and (d) of
Fig.~2 for the time steps $h^\ast=0.00125$, 0.0025, 0.005, and
0.01, respectively. For the purpose of comparison, the same
dependencies corresponding to original VFR and PFR integrators
(see Eq.~(8)) are also shown there. Note that for the original
integrators, the time step for each subset was chosen to be always
in factor 4/3 smaller than that of the optimized versions, namely,
$h^\ast=0.0009375$, 0.001875, 0.00375, and 0.0075. The last
condition is necessary to provide the same number of total force
recalculations during the same observation time $t \gg h$ within
both the approaches. Note also that within these approaches, the
fifth- and higher-orders truncation uncertainties become too big
at step sizes $h^\ast
> 0.01$. In particular, then the ratio of the total energy fluctuations to
the fluctuations in potential energy (the standard quantity to estimate the
precision of the calculations) exceeds a few per cent. Such large step sizes
cannot be used in precise MD simulations and, for this reason, are not
considered here.

\begin{figure}[htbp]
\begin{centering}
\begin{picture}(84,200)
\epsfxsize=84mm \put(0,0){ \epsffile[27 280 558 702]{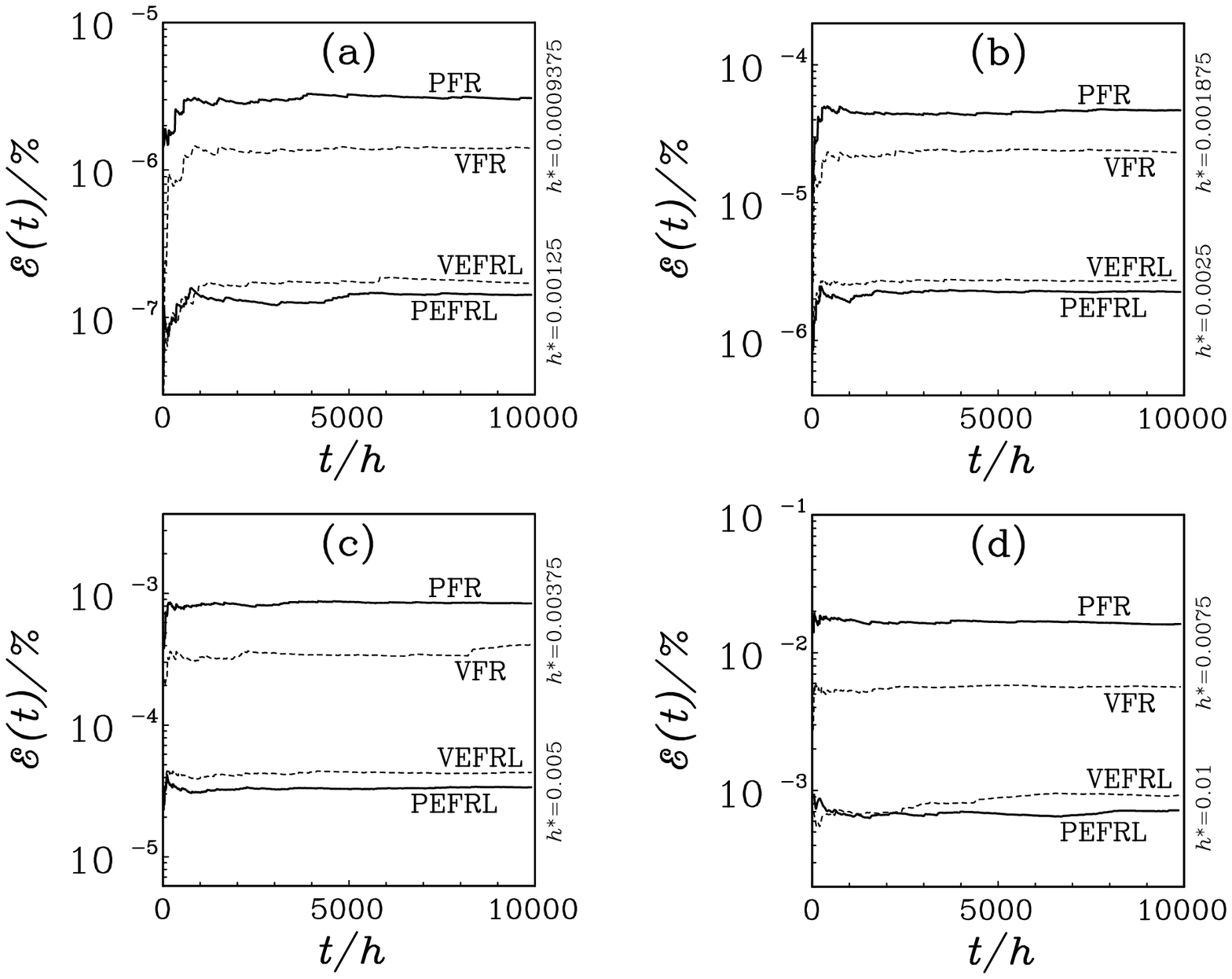}}
\end{picture}
\end{centering}
\end{figure}
{\small FIG. 2. The total energy fluctuations as functions of the
length of the simulations performed using the optimized VEFRL and
PEFRL algorithms, as well as the original VFR and PFR integrators.
The results corresponding to different values of the time step,
namely, $h^\ast=0.00125$ and 0.0009375, 0.0025 and 0.001875, 0.005
and 0.00375, as well as 0.01 and 0.0075 are presented in subsets
(a), (b), (c), and (d), respectively.}

\vspace{7pt}

As can be seen from Fig.~2, both the original (VFR and PFR) and
optimized (VEFRL and PEFRL) algorithms exhibit very good stability
properties, that is explained by the symplecticity and time
reversibility of the produced solutions. No systematic deviations
in the total energy fluctuations can be observed for all the
integrators. Instead, in each of the cases the amplitude of the
fluctuations quickly tends to its own value which does not
increase with further increasing the length of the simulations.
However, this value is significantly larger within the original
VFR and PFR integration. At the same time, using the optimized
VEFRL and PEFRL algorithms allows us to decrease the unphysical
energy fluctuations approximately in factor 10--15 (for VEFRL) or
20--25 (for PEFRL), despite the larger time steps. This is in
excellent accord with our theoretical predictions $\Gamma_{\rm
min}^{\rm VEFRL}(4h/3)/\Gamma_{\rm FR}(h) \approx 0.073$ and
$\Gamma_{\rm min}^{\rm PEFRL}(4h/3)/\Gamma_{\rm FR}(h) \approx
0.051$, obtained in section III. It is worth pointing out also
that the PEFRL algorithm is slightly better in energy conservation
than its VEFRL counterpart (whereas the VFR integrator is better
with respect to its PFR version).

Consider now the results for composition integration. The total
energy fluctuations obtained in this case at the end of the
simulation runs for the same set (as in Fig.~1) of undimensional
time steps, $h^\ast = h (u/m\sigma^2)^{1/2}=0.00125$, 0.0025, and
0.005, are shown in Fig.~3 as depending on parameters $\xi$
(subsets (a) and (c)) and $\chi$ ((b) and (d)). The subsets
(a)--(b) and (c)--(d) relate to the VESL and PESL versions,
respectively, of composition scheme (25). Mention that the time
coefficients $\xi(\chi)$ and $\lambda(\chi)$ for this scheme
should satisfy equality (31) in which the quantity $\chi=1 - 2
(\xi+\lambda)$ is treated as a free parameter. So that only one
$\chi$-dependency is enough, in principle, to present the results.
But, in view of somewhat complicated structure of function ${\cal
E}(\chi)$, we have included the $\xi$-dependency as well for more
clarity. Note also that according to constraint (31), the
parameter $\chi$ cannot accept values from the interval
$]1-4\vartheta \approx -0.658, 1]$, because otherwise this will
lead to imaginary solutions for $\xi$ and $\lambda$.

\begin{figure}[htbp]
\begin{centering}
\begin{picture}(84,195)
\epsfxsize=84mm \put(0,0){ \epsffile[41 288 554 702]{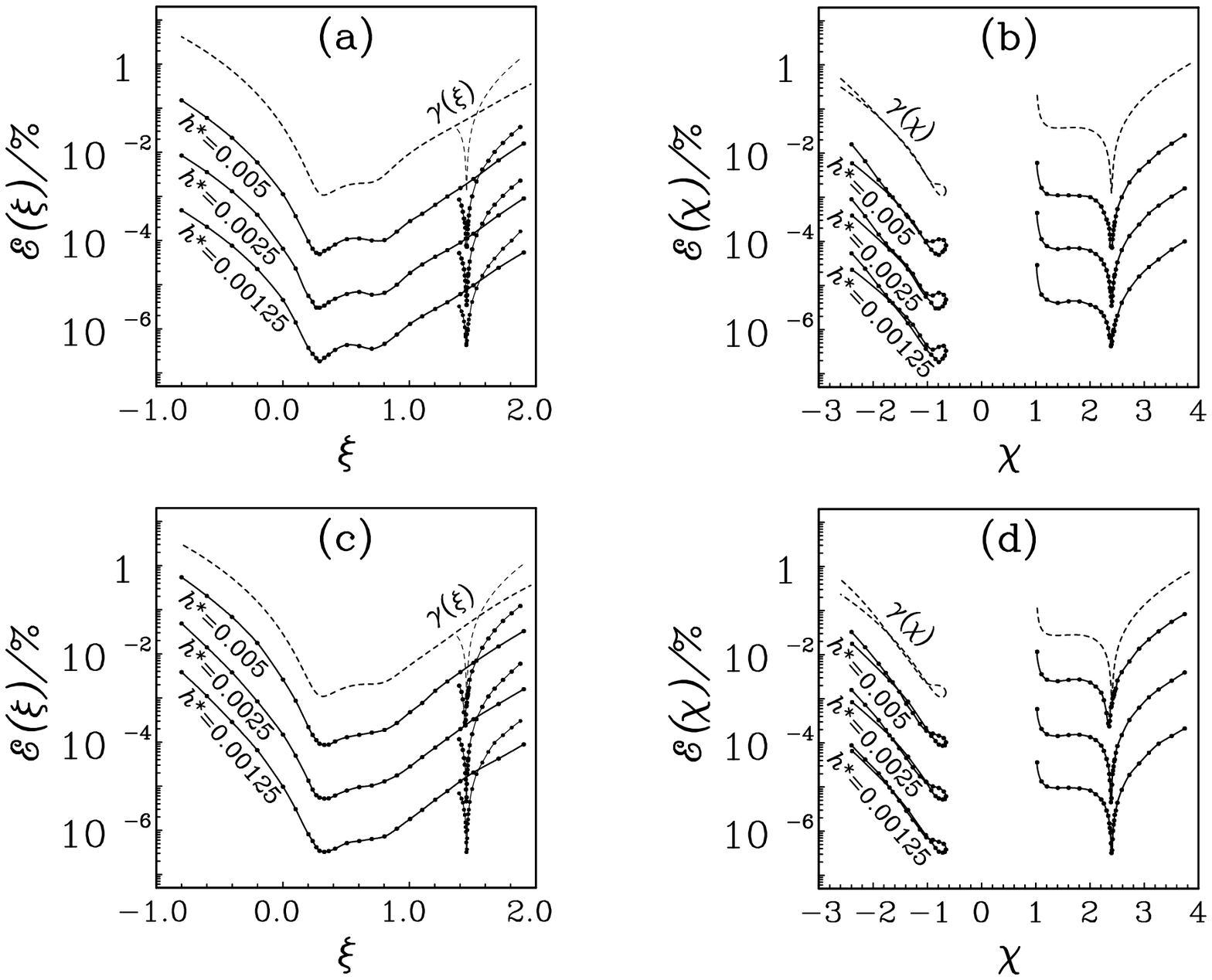}}
\end{picture}
\end{centering}
\end{figure}
{\small
FIG. 3. The total energy fluctuations obtained in the simulations for
different values of parameters $\xi$ and $\chi$ at three fixed time
steps, $h^\ast=0.00125$, 0.0025, and 0.005 using the VESL (subsets (a)
and (b)) and PESL ((c) and (d)) versions of composition integration
(Eq.~(25)). The simulation results are presented by circles connected
by the solid curves. The function $\gamma$ (see Eq.~(29)) are plotted
by the dashed curves.}

\vspace{7pt}

As can easily be seen from Fig.~(3), all the dependencies have up three
minima and their locations on $\xi$- and $\chi$- axes practically do not
depend on the size $h$ of the time step. Moreover, the fluctuations
${\cal E}(\xi,h)$ again appear to be proportional to the norm $\Gamma=
\gamma h^4$ of global errors, where the function $\gamma$ is now defined
by Eq.~(29) and plotted in the figure as well. Among the three minima,
two of them have nearly the same depth, and for VESL integration the
global minimum of ${\cal E}$ is achieved at $\xi \approx 0.323$ or $\chi
\approx -.726$, i.e., from the left of the interval $]1-4\vartheta, 1]$
for $\chi$. This coincides with the predicted values given by Eq.~(35)
corresponding to the global minimum of $\gamma$. In the case of PESL
integration the pattern is somewhat different. At small enough values
of the time step, namely at $h \le 0.0025$, the global minimum for the
functions ${\cal E}$ and $\gamma$ is identified from the right of the
above $\chi$-internal, namely, at $\xi \approx 1.45$ or $\chi \approx
2.4$. But with further increasing $h$, the position of this minimum for
${\cal E}$ begin to shift with respect to its initial value. Such a
behavior can be explained by the influence of higher-order ${\cal O}(h^7)$
truncation errors, which appears to be significant in this case due to
the largeness of absolute values of time coefficients $\xi$, $\lambda$,
and $\chi$ (in particular, then all these coefficients are larger in
amplitude than unity). This is contrary to a left-lying minimum of
${\cal E}$ which is achieved at the same point $\xi \approx 0.316$ or
$\chi \approx -0.737$ as that of function $\gamma$ (see Eq.~((34)
independently on the size of the time steps considered. Then taking
into account that both the minima have almost the same depth, and the
fact that the left-lying minimum of ${\cal E}$, being local at small
$h$, transforms into the global minimum at moderate and larger time
steps, solutions given by Eq.~(34) should be considered as optimal.

As expected, at the minima found the energy fluctuations decrease
approximately in 1.5 times with respect to those at $\xi = \lambda
\equiv \vartheta \approx 0.41$ (and $\chi \approx -0.66$) corresponding
to the original Suzuki approach. It is interesting to point out that the
constraint $\xi = \lambda$ minimizes the function $\sum_{q=1}^5 |d_q|
\equiv 2|\xi|+2|\lambda|+|\chi|$ and provides a minimum for the maximal
value among the quantities $|\xi|$, $|\lambda|$, and $|\chi|$. Using
these last two criteria and the additional requirement that $|d_q| < 1$,
Kahan and Li \cite{KahanLi} have derived extended composition schemes like
(23) up to the tenth order in the time step. In view of our more precise
analysis performed in this paper we see, therefore, that the previous
criteria for optimization of algorithms may have nothing to do with
providing the best performance of the computations. Finally, comparing
the results presented in Figs.~1 and 3 between themselves, it can be
pointed out that the values for minima of ${\cal E}$ corresponding to
the VESL and PESL integration are approximately in 1.5 times higher
than those of the VEFRL and PEFRL algorithms. In addition, VESL and PESL
integrators require more (five, instead of four) force evaluations per
time step, so that their efficiency is in factor 1.5 $(5/4)^4 \approx 3$
worse with respect to the VEFRL and PEFRL schemes. Therefore, as was
predicted (see. Eq.~(33)), the preference should be given to the
decomposition VEFRL and PEFRL algorithms rather than to their
composition counterparts VESL and PESL.

\section{Conclusion}

We have formulated a new approach for improving the efficiency of
decomposition integration of the equations of motion in many-body
systems. As a result, the Forest-Ruth-like algorithms have been derived
and optimized for MD as well as celestial mechanics simulations. The
algorithms are explicit, simple in implementation, and produce phase
trajectories which are exactly symplectic and time reversible. Their
main advantage over the original integrators by Forest and Ruth is the
possibility to generate much more precise solutions at the same overall
computational efforts. It has been predicted theoretically and confirmed
in actual MD simulations of a LJ fluid that the optimized Forest-Ruth-like
algorithms allow to decrease significantly the unphysical fluctuations of
the total energy using even greater time steps. The question of how to
optimize the integration by composing second-order schemes has also been
studied. As has been shown, the Suzuki-like algorithms, obtained in this
case, although improve the efficiency of the computations with respect to
the original integrators by Suzuki as well as Forest and Ruth, appear to
be less efficient than the optimized Forest-Ruth-like algorithms. The
last algorithms should be considered as optimal among all possible
decomposition integrators of the fourth-order with single splitting
of the Liouville operator.

The approach presented can also be adapted for optimization of the
integration of motion in more complicated systems for which splitting
of the Liouville operator into more than two parts may pay dividends.
The examples are multi-component fluids and systems with long-range
interactions, where characteristic time intervals of the dynamical
processes can differ by many orders from each other. In some cases,
for instance, for systems with orientational degrees of freedom, the
additional splitting may be necessary to obtain analytically integrable
parts. These and related problems will be considered in a separate
publication.

\vspace{4pt}

Part of this work was supported by the Fonds zur F\"orderung der
wissenschaftlichen Forschung under Project No. P15247.

\begin{center}
{\bf
Appendix. Optimization of Forest-Ruth-like algorithms
          in specific cases}
\end{center}
\setcounter{equation}{0}
\renewcommand{\theequation}{A\arabic{equation}}

In section III, we have derived the velocity- and position-Forest-Ruth-like
algorithms and optimized them for integration of motion in MD simulations.
Now, one considers two specific cases when the efficiency of decomposition
scheme (9) can be improved additionally. One example is a collection of
weakly interacting particles, in which the potential part $B \equiv
\varepsilon {\cal B}$ of the Liouville operator can be treated as a small
perturbation with respect to its kinetic counterpart $A$, where $\varepsilon
\ll 1$. Another case is the integration of motion in the solar system, where
the interactions between planets are negligible small in comparison with
the force that acts on them with respect to the sun. Then, the Liouville
operator can be presented as $L = {\cal A} + \varepsilon {\cal B}$, where
now ${\cal A}$ is the sum of two-bodies Liouville's commutative operators,
each of which describes the motion of an isolated body in an external field
of the central mass, whereas $\varepsilon {\cal B}$ corresponds to weak
interactions of the bodies between themselves. Here, as in the usual case,
the separate exponential operators ${\rm e}^{{\cal A} \tau}$ and ${\rm
e}^{{\cal B} \tau}$ appear to be exactly integrable, so that the
decomposition scheme will also lead to explicit integration.

Therefore, assuming that the parameter $\varepsilon$ is sufficiently
small and considering the velocity version of extended Forest-Ruth-like
decomposition (9), we can neglect by fifth-order commutators arising in
the truncation term $C_5$ at the multipliers $\gamma_2$, $\gamma_3$,
$\gamma_4$, $\gamma_5$, and $\gamma_6$ (see Eq.~(12)). The reason is that
such commutators are of order $\varepsilon^2$ or higher, whereas the
commutator at $\gamma_1$ is proportional only to the first power of
$\varepsilon$. In such a case, it is quite natural to reduce the
multiplier $\gamma_1$ to zero exactly, rather than to find a minimum
for the norm (13) of $\gamma$ with respect to all the commutators. Then
the problem (15) transforms into the following system of three equations
\begin{equation}
\left \{
\begin{array}{l}
\alpha(\xi,\lambda,\chi)=0 \, , \\ [3pt]
\beta(\xi,\lambda,\chi)=0 \, ,  \\  [3pt]
\!\! \gamma_1(\xi,\lambda,\chi)=0 \, ,
\end{array}
\right.
\end{equation}
where $\alpha$ and $\beta$ are defined according to Eq.~(11), and the
expression for $\gamma_1$ follows immediately after Eq.~(12). Since there
exist up four sets of real solutions to Eq.~(A1), an additional requirement
should be imposed to choose the optimal values among them. Obviously, such
values must minimize the norm of the main term of truncated uncertainties,
which now is equal to $\sqrt{\gamma_2^2+\gamma_3^2}$ (note that commutators
at $\gamma_2$ and $\gamma_3$ are proportional to $\varepsilon^2$, while to
the third power of $\varepsilon$ or higher at $\gamma_4$, $\gamma_5$, and
$\gamma_6$). As a result, one finds the optimal solutions
\begin{eqnarray}
\xi     =  \ \  &&0.5316386245813512  \nonumber \\
\lambda =  \ \  &&0.5437514219173741            \\
\chi \! =      -&&0.3086019704406066  \, . \nonumber
\end{eqnarray}

For the position version (when $A \leftrightarrow B$), the commutators
at $\gamma_1$ and $\gamma_4$ are interchanged, so that it is necessary
to solve the system $\alpha=0$, $\beta=0$, and $\gamma_4=0$. No real
solutions to this system have been found. In such a situation, we could
try to find the global minimum for $|\gamma_4|$. But since it appears
to be greater than zero, the position version is not recommended to be
used for this case and the preference should be given to the extended
integration in velocity form (with time coefficients defined by Eq.~(A2)).
Note that the main terms of uncertainties for this integration are ${\cal
O}(\varepsilon^2 h^5)$ and ${\cal O}(\varepsilon h^7)$, so that in view
of the smallness of $\varepsilon$, the scheme derived can be related
to a quasi sixth-order algorithm. Such an algorithm, contrary to
usual sixth-order schemes, will require four, instead of seven
\cite{Yoshida,Suzukium}, force evaluations per time step.

In the above derivation, we tentatively assumed that the order of
smallness of $\varepsilon$ is nearly the same as that of the size for
the time step $h$. In systems for which the parameter $\varepsilon$ is
extremely small, the strategy of optimization should be reconstructed.
Namely, we can now ignore all the terms of the second order with respect
to $\varepsilon$, even that which appears at $\beta$ (see Eq.~(10)) in
third-order uncertainties ${\cal O}(h^3)$ with respect to $h$. Then
within the velocity version, one comes to two equations, $\alpha=0$
and $\gamma_1=0$, for three variables, $\xi$, $\lambda$, and $\chi$.
Thus, the order of the decomposition scheme can be increased additionally
by canceling an extra-order term ${\cal O}(\varepsilon h^7)$ (which is
assumed to be much more significant than the terms ${\cal O}(\varepsilon^2
h^3)$ and ${\cal O}(\varepsilon^2 h^5)$). An explicit expression for
${\cal O}(\varepsilon h^7)$ (it is denoted simply as ${\cal O}(h^7)$ in
decomposition formula (9)) has the form
$$
{\cal O}(\varepsilon h^7) = \big( c_7^{(A)} [A,[A,[A,[A,[A,[A,B]]]]]] +
{\cal O}(\varepsilon^2) \big) h^7 \, ,
$$
where
$$
c_7^{(A)}=  -\frac{31}{967680} + \lambda^2 \chi \bigg( \frac{7}{5760} -
\frac{\lambda^2}{288} + \frac{\lambda^4}{360} \bigg) + \frac{\xi}{7680}
\, .
$$
So that the desired system of equations is
\begin{equation}
\left \{
\begin{array}{r}
\alpha(\xi,\lambda,\chi)=0 \\ [3pt]
\!\! \gamma_1(\xi,\lambda,\chi)=0 \\ [3pt]
c_7^{(A)}(\xi,\lambda,\chi)=0
\end{array}
\right.
\end{equation}
It can be solved analytically, and the solutions are
\begin{equation}
\xi     = \frac{1}{20} \, , \ \ \ \ \ \ \
\lambda = \pm \frac{1}{2} \sqrt{\frac{3}{7}} \, , \ \ \ \ \ \ \
\chi    = \frac{49}{180} \, ,
\end{equation}
where the preference should be given for sign ``$+$'', because this
leads to a smaller value ($\approx 0.0036$) for $|\beta|$ (and for an
multiplier at the ninth-order commutator) which forms the main term
${\cal O}(\varepsilon^2 h^3)$ (and ${\cal O}(\varepsilon h^9)$) of
local truncation errors.

In the case of position version ($A \leftrightarrow B$), the seventh-
order (with respect to $h$) truncation term behaves as
$$
{\cal O}(\varepsilon h^7) = \big( c_7^{(B)} [B,[B,[B,[B,[B,[A,B]]]]]] +
{\cal O}(\varepsilon^2) \big) h^7 \, ,
$$
with
\begin{eqnarray*}
c_7^{(B)}=&&-\frac{1}{30240} + \lambda \chi \bigg( \chi \bigg[\frac{1}{720} -
\frac{\xi^2}{24}+\frac{\xi^3}{12}-\frac{\xi^4}{24}\bigg] -
\nonumber \\ &&\!
\chi^2 \bigg[\frac{\xi}{36} - \frac{\xi^2}{12} + \frac{\xi^3}{18}\bigg] -
\chi^3 \bigg[\frac{1}{144} - \frac{\xi}{24} + \frac{\xi^2}{24}\bigg] +
\nonumber \\ &&\!
\chi^4 \bigg[\frac{1}{120} - \frac{\xi}{60}\bigg] - \frac{\chi^5}{360} +
\frac{\xi}{360} - \frac{\xi^3}{36} + \frac{\xi^4}{24} -
\frac{\xi^5}{60} \bigg) +
\nonumber \\ &&\!
\ \ \ \ \
\frac{\xi^2}{1440} - \frac{\xi^4}{288} +
\frac{\xi^5}{240} - \frac{\xi^6}{720}
\nonumber
\end{eqnarray*}
and equations (A3) transforms into
\begin{equation}
\left \{
\begin{array}{r}
\beta(\xi,\lambda,\chi)=0 \\ [3pt]
\!\! \gamma_4(\xi,\lambda,\chi)=0 \\ [3pt]
c_7^{(B)}(\xi,\lambda,\chi)=0
\end{array}
\right.
\end{equation}
System (A5) has eight solutions and all of them are real. The optimal
solution, which minimizes (to the value $\approx 0.0034$) the multiplier
$|\alpha|$ (it forms now the main term ${\cal O}(\varepsilon^2 h^3)$ of
local uncertainties), is
\begin{eqnarray}
\xi    &=& \frac{1}{2} - \frac{1}{70} \sqrt{525 + 70 \sqrt{30}}
\nonumber \\ [3pt]
\lambda&=& \frac{1}{4} + \frac{\sqrt{30}}{72}
\\ [3pt]
\chi   &=& \frac{1}{\sqrt{70}} \sqrt{15 - \sqrt{105}}
\nonumber
\end{eqnarray}
For both velocity (A4) and position (A6) versions, the main terms of
uncertainties are ${\cal O}(\varepsilon^2 h^3)$ and ${\cal O}
(\varepsilon h^9)$. So that, taking into account the smallness of
$\varepsilon$, these versions present, in fact, quasi eight-order
algorithms, which contrary to usual eight-order schemes, will require
again only four, instead of up fifteen \cite{Yoshida,Suzukium}, force
evaluations per time step.

\end{multicols}
\end{document}